\documentclass[sigplan,10pt,authorversion=true]{acmart}
\AtBeginDocument{%
  \providecommand\BibTeX{{%
    \normalfont B\kern-0.5em{\scshape i\kern-0.25em b}\kern-0.8em\TeX}}}

\copyrightyear{2021} 
\acmYear{2021} 
\setcopyright{acmlicensed}\acmConference[PLOS '21]{11th Workshop on Programming Languages and Operating Systems}{October 25, 2021}{Virtual Event, Germany}
\acmBooktitle{11th Workshop on Programming Languages and Operating Systems (PLOS '21), October 25, 2021, Virtual Event, Germany}
\acmPrice{15.00}
\acmDOI{10.1145/3477113.3487274}
\acmISBN{978-1-4503-8707-1/21/10}

\begin{document}

\title{How ISO C became unusable for operating systems development}

\author{Victor Yodaiken}
\email{vy@e27182.com}
\orcid{1234-5678-9012}
\affiliation{%
  \institution{E27182}
  \streetaddress{}
  \city{Austin}
  \state{Texas}
  \country{USA}
  \postcode{78733}
}

\renewcommand{\shortauthors}{Yodaiken}

\begin{abstract}
The C programming language was developed in the 1970s as a fairly unconventional systems and operating systems development tool, but has, through the course of the ISO Standards process, added many
attributes of more conventional 
programming languages and
become less suitable for operating systems development. 
Operating system programming continues to be done in non-ISO dialects of C. The differences provide a glimpse of operating system requirements for programming languages. 
\end{abstract}




\begin{teaserfigure}
 \includegraphics[width=\textwidth,height=7cm]{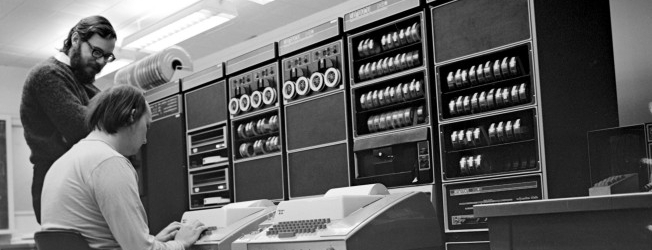}
  \caption{Ken Thompson and Dennis Ritchie at Bell Labs}
  \Description{Ken Thompson and Dennis Ritchie at Bell Labs}
  \label{fig:dmkt}
\end{teaserfigure}

\maketitle

\section{Introduction}
The C programming language \cite{ritchie1978c} is the first, and, so far, only widely successful programming language that provides operating system developers with a high-level language alternative to assembler (compare to \cite{bliss}).  C's success was predicated
on its design: a small language, close to the machine yet with a great deal of flexibility for experienced programmers. The Rationale for the C standard \cite{rationalec} cited C's capability to function as a "high-level assembler" and 
explained that \emph{"many operations are defined to be how the target machine’s hardware does
it rather than by a general abstract rule"} but C also has traditional attributes of an ALGOL style programming language. 

At present most major commercial operating system kernels and many experimental ones are written primarily in C or some dialect.  However,
ISO C \cite{ANSIC89,c18}, the language that has evolved over nearly forty years of the standards process,
has not only diverged from Kernighan and Ritchie C (K\&R C) \cite{kandr}, but has become poorly suited to operating systems development.
C based operating systems projects on ISO C-compliant compilers \cite{gcc,clang} rely on compiler specific opt-outs, assembler escapes, and coding tricks
to produce a usable C dialect, largely undermining the purpose of an ISO standard.  Among the techniques used in Linux are multiple \textsf{gcc}-specific opt-outs (\cite{wang} fig. 9) and "opaque" operations such as in-line assembler code that 
hides some pointer addition for per-CPU data structures \cite{relochide}. These techniques
are not necessarily stable or reliable as the compilers evolve and change C semantics.

A common argument (made e.g.\ by Dietz \cite{dietz}) is that
programmers are wrong: their objections to changes in C semantics embody  "a fundamental and pervasive misunderstanding: the compiler [is] not 'reinterpreting' the semantics but
rather [is] beginning to take advantage of leeway explicitly provided by the C standard." However, that overlooks the rationale's claimed intention to continue supporting the high-level assembler use case.

Limitations of ISO C for OS development have been noted in academic literature 
\begin{quote}"\emph{Systems or library C codes often cannot be written in standard-conformant C}" \cite{compcert2}.\end{quote}
and by practitioners e.g.  \cite{torvaldsalias}. 
The primary cause is a design approach in the ISO standard that has given priority to certain kinds of optimization over
both correctness and the "high-level assembler" \cite{rationalec} intentions of C, even while the latter remain enshrined in the rationale.

For example, a well-known security
issue in the Linux kernel was produced by a compiler 
incorrectly assuming a pointer null check was unnecessary (\cite{wang} fig. 6) and deleting it as an optimization.
Or consider this (simplified) patch report for
Linux \cite{nossum}:
\begin{quote}
    The test  [for termination] in this loop: [...]
was getting completely compiled out by my \textsf{gcc}, 7.0.0 20160520. The result
was that the loop was going beyond the end of the [...] array and
giving me a page fault  [...]

I strongly suspect it's because \verb|__start_fw| and \verb| __end_fw|
are both declared as (separate) arrays, and so \textsf{gcc} concludes that \verb|__start_fw| can
never point to \verb|__end_fw|.

By changing these variables from arrays to pointers, \textsf{gcc} can no longer
assume that these are separate arrays.
\end{quote}
Here, \textsf{gcc} is "optimizing" based on the assumption that one externally defined object cannot overlap another, even though the host platform allows this.

ISO delegates to the compiler a great deal of the control that K\&R C divides between the programmer, the environment, and the
target architecture but \emph{not} the compiler, as witnessed by the rationale's reference to ``how the target machine does it''.
Consider a simple scalar read \verb|*x = y|. In K\&R C, a programmer could be reasonably certain that code would be generated  to copy a value from the storage starting at the  location  of \verb|y| to the location with address \verb|x|, something like \verb|load x,Reg1;|
\verb|load (y),Reg2;| \verb|store Reg2,(Reg1)| or perhaps simpler if some values are already cached in registers.
In ISO C, things are complicated. The compiler might generate no code at all, if, for example it
detects the address in \verb|x| to be aliasing a storage location that is used elsewhere with a different enough type. Or
the compiler might conclude that \verb|y| is uninitialized and generate code to copy some arbitrary value to \verb|*x| on the assumption that any arbitrary value will do, perhaps a different one on each access. The compiler is often free to reorder statements if it detects or assumes they do not have data dependencies. 

For an example of an implementation of \texttt{malloc} in K\&R (page 187) the text explains there is a question about whether
"pointers to different blocks ... can be meaningfully compared", something not guaranteed by the standard. The conclusion is "this version of \texttt{malloc} is portable only among machines for which general pointer
comparison is meaningful." -- delegating the semantics to the processor architecture.
There is no suggestion that the code is invalid. In contrast,
consider the much higher-level, compiler-controlled view of comparing pointer equality in \cite{provenance} 
(my bold).
\begin{quote}A priori, pointer equality comparison (with == or != ) might be
expected to just compare their numeric addresses, but we observe \textsf{gcc} 8.1 -O2 sometimes regarding two pointers
with the same address but different provenance as nonequal. Unsurprisingly, this happens in some circumstances
but not others, e.g.\ if the test is pulled into a
simple separate function, but not if in a separate
compilation unit. To be conservative w.r.t.\ current compiler behaviour,
pointer equality in the
semantics should give false if the addresses are
not equal, but \textbf{nondeterministically} (at each run-time occurrence)
either take provenance into account or not if the addresses are equal – this specification
looseness accommodating implementation variation.
\end{quote}

The ISO approach of delegating to the compiler, not the machine, has damaging effects on
reliability and expressive range for OS programming.
The next two sections discuss further examples of this.
Since compiler optimizations are the reason behind this approach,
the final section discusses reasons for skepticism about whether this is necessary or helpful
in this specific domain.  Whether these kinds of optimizations
are necessary for other types of applications is not within scope.

\paragraph{Related Work and Scope}
There have been a number of articles and essays on the controversial  "undefined
behavior" mechanism employed to enable many ISO C optimizations
( e.g., \cite{wang,wang2,regehrguide,ertlwhat,lattner}): the effects on operating system 
programming are discussed in section \ref{sec:ub}. 
The complexity of ISO C semantics has also motivated development of formalizations  intended to be easier to reason about, more precisely specified, or more useful for low-level programming \cite{kang,hathorn,compcert2} and extensions to compiler intermediate languages \emph{e.g.} Lee \cite{lee}. 
These are clearly related works, as discussed below, but they assume exactly what is being questioned here: the utility of the optimization approach of ISO C. 
The CompCert compiler \cite{Leroy-Compcert-CACM} is discussed in the final section. 

\section{Optimization and time bombs}
Dennis Ritchie \cite{dmr} wrote the following as part of an objection to  one of the first ANSI C standard drafts: 
\begin{quote}
    The fundamental problem is that it is not possible to write real programs using the X3J11 definition of C. The committee has created an unreal language that no one can or will actually use.
\end{quote}
Ritchie's main objection was to a type attribute intended to limit aliasing (two or more active pointers/references
addressing the same storage). 
\begin{quote}
the committee is planting timebombs that are sure to explode in people’s faces. Assigning an ordinary pointer to a pointer to a `noalias’ object is a license for the compiler to undertake aggressive optimizations that are completely legal by the committee’s rules, but make hash of apparently safe programs. 
\end{quote}
C's pointer system can be a problem for optimizing compilers. 
Even for something as apparently innocent as 
\begin{verbatim}for(i=0; i < *b; i++)a[i] = a[i]+ *v; \end{verbatim}
if it is possible that for some \verb|i|, \verb|a+i == v| or \verb|a+i == b| or even \verb|a+i = &v| so
compiled code must reload both values on each iteration of the loop\footnote{Here I am treating pointer equality in terms of machine addresses.}.  
In theory, the more the compiler can restrict aliasing, the more it can optimize. Stronger alias rules should permit more common subexpression elimination and redundancy elimination.  Ritchie was dubious:
\begin{quote}
 Perhaps there is some reason to provide a mechanism for asserting, in a particular patch of code, that the compiler is free to make optimistic assumptions about the kinds of aliasing that can occur. 
 I don't know any acceptable way of changing the language specification to express the possibility of this kind of optimization, and I don't know how much performance improvement is likely to result. 
 \end{quote}
The ANSI Committee, soon to become the ISO Committee,  backed down in the face of Ritchie's objections --- temporarily.  A year later, C89 imposed type restrictions on access to C objects as a way
of facilitating type-based alias analysis (TBAA) (section 3.3 in C89). 
The basic idea is that ISO C forbade accessing an object of one type via a pointer (or other "left hand side") of a different enough type,  with an exception for character pointers which can access everything (sort of -- as discussed below).

The usual example  of TBAA optimization 
 involves "lifting" variables out of loops. For the loop above 
 \begin{verbatim}
long  b1 = *b; long  v1 = *v; //lifted
for(int i= 0; i < b1; i++)a[i]= a[i]+v1); \end{verbatim}
 is a legal optimization if the type of \verb|a[i]| doesn't match the types of \verb|*b| and \verb|*v| because the compiler "knows" that those pointers cannot alias objects of a different type.

 The standard does not require compilers to flag aliasing violations or to prove the absence of aliasing. 
 Instead the ISO standard permits compilers to \emph{assume} an absence of aliasing in a number of cases including cases
 where there would otherwise be type violations. 
Consider the following  code fragment \cite{aliasfloat} where a floating point value is set to \(-3.14\) and then an aliasing unsigned int
pointer is used to turn off the sign bit. With no optimization, under \textsf{Clang} 12.01, this program prints \(3.14\).
With optimization level 2, it prints \(-3.14\) because,
assuming the pointer cannot alias the floating point variable, the compiler discards the mask operation as an optimization.
\begin{verbatim}
  float  f; long *l = (long *)&f;
  f= -3.14;
  //forbidden aliasing
  *l  &= 0x7fffffff;
  printf("f = %f \n", f); return;
\end{verbatim}
 Since the  assumption that there is no aliasing is false in this case, a programming type error is silently "optimized" to
 a logic error.
 
Under these rules, radix sort is only permitted for elements that have byte-sized radixes.

 \subsection{Effects on Operating System development}
For operating systems the effects are widespread particularly for objects that have different semantics depending on which 
kernel component is accessing them. 
A single block of storage may be addressed
by the disk manager as a block of unsigned characters, by the file manager as an array of inodes, a directory block, or 
a block of characters, 
and by a page manager via a void pointer, all at the same time. The aliasing rules of ISO C are not 
compatible with this approach. It may be possible in ISO C to push all these different types into a union, but
that would harm modularity, by requiring each of the components to share the basic data structures of the others. 

As another example,
computing a checksum for a data structure
by aliasing it with an int pointer is not permitted.
\begin{verbatim}
packet_t * p = getpacket();
int ck=0;
int i;
int *q = (int *)p;  //cast is ok
for(i= 0; i< sizeof(packet_t)/sizeof(int); i++)
            ck ^= q[i]; //not permitted 
\end{verbatim}

There is no general escape mechanism, even though char pointers are allowed to alias anything\footnote{There is a \textsf{Clang}-specific "may\_alias" attribute and \texttt{memcpy} is sometimes suggested as a work-around but it introduces semantically confusing additional copying of data with hope that the optimizer may be able to do what the programmer could not do directly.}. 
The absence of escapes  is a major change in C's type system as Brian Kernighan's critique of Pascal makes clear \cite{kernighan}: 
    \begin{quote}\textit{
        There is no way [in Pascal] to override the type mechanism when necessary, nothing analogous to the ``cast'' mechanism in C.  This means that it is not possible to write programs like storage allocators or I/O systems in Pascal, because there is no way to talk about the type of object that they return, and no way to force such objects into an arbitrary type for another use.}  
    \end{quote}
As we saw earlier, the implementation of \texttt{malloc}/\texttt{free} in K\&R is not conformant ISO C code. ISO C needs special rules around ``effective types'' (explored later) in  order to permit \texttt{malloc}/\texttt{free} to work. Even then, it remains unclear how to write these functions in conformant ISO C.  

Perhaps the most important effect is the loss of semantic clarity.
Programmers are basically mystified
by the rules: see  "The Strict Aliasing Situation is Pretty Bad"  \cite{regehralias} (the comments are especially illuminating). 
Thirty years after the aliasing type restrictions went into the Standard, a committee of standards experts wrote that the 
current situation "leaves
many specific questions unclear: it is ambiguous whether some programming idioms are allowed or not, and exactly
what compiler alias analysis and optimisation are allowed to do." \cite{provenance}.

 In practice, alias analysis in \textsf{gcc} and \textsf{Clang} has unpredictable effects. \textsf{Gcc} does not omit the mask operation in the 
 floating point example above although it does do typed alias optimizations sometimes (see Figure 7 in \cite{wang} for an example).
 Linux uses a flag to disable "strict-aliasing" analysis in \textsf{gcc} \cite{torvaldsalias2,torvaldsalias}
 but that does not disable all alias optimizations as shown by an example in \cite{provenance} (page 15).
And because aliasing violations may not be flagged, there can be silent, surprising, changes in code
operation between optimization levels or versions.
This is not a C-specific problem. Fortran has similar problems with similar assumptions: 
 "anything may happen:
the program may appear to run normally, or produce incorrect answers, or behave
unpredictably." \cite{nguyenalias}.
 
Violation of aliasing rules is just one example of a large class of "undefined behaviors" and the next section looks at that
topic more generally. 

\section{Undefined behavior and land mines}\label{sec:ub}
Neither K\&R2 nor \cite{ritchie1978c} mentions "undefined behavior", but it is a central if controversial 
concept in ISO C. Good summaries can be found in \cite{regehrguide,wang,hathorn, ertlwhat, lattner}. 
As described in the C "Rationale" \cite{rationalec}, undefined behavior is a modest concept:
\begin{quote}
    Undefined behavior gives the implementor license not to catch certain program errors that are
difficult to diagnose. It also identifies areas of possible conforming language extension: the
implementor may augment the language by providing a definition of the officially undefined
behavior. 
\end{quote}
This is a relatively simple, maybe deceptively simple, idea that could be interpreted, for example, as permitting a C implementation to use single
machine instructions for basic arithmetic operations and let the hardware handle (or ignore) arithmetic overflow. The C standard has
long declared signed integer overflow to be undefined behavior and this interpretation would permit the wrapping behavior native to most modern processors,  the saturating arithmetic of some controllers, or the more widely varied behaviors of historical processors -- all implemented efficiently by adopting the target processor-native semantics. This modest view of undefined behavior is not, however, the prevailing one, which is that the compiler can assume undefined behavior is impossible and can optimize on the basis of that assumption. In fact,
it is currently argued that the standard interpretation allows implementations to take any action at all, not just for (say) an overflowing execution but for the entire program, if they detect a single
feasible instance of undefined behavior. And there are lots of undefined behaviors.

By C18, the ISO C Standard document included a 10-page, incomplete list of undefined behaviors covering everything from type constraints to  syntax errors and synchronization errors. Most C programs contain undefined behavior -- certainly every operating system code base does. 
Perhaps more troubling, as \cite{ertlwhat} points out, this concept of undefined behavior makes C compilers unstable. A programmer may take a particular
 property of a C compiler for some undefined behavior to be a conforming language extension, but it may actually just be undefined behavior that has not yet been optimized. For example, \textsf{gcc} will generally ignore type rules on pointers - except when it does not - so that \verb|*p = k| may work at one level of optimization, perhaps for decades, but be deleted silently when the optimizer pass recognizes a type mismatch.  Kang \cite{kang} notes the "somewhat controversial practice of sophisticated C compilers reasoning backwards
from instances of undefined behavior to conclude that, for example, certain code paths must be dead." can  lead to "surprising non-local changes in program behavior and difficult-to-find bugs".
 
And by 2011, Chris Lattner, the main architect of the \textsf{Clang}/LLVM compilers was echoing Ritchie's warning \cite{lattner2}: 
\begin{quote}
    To me, this is deeply dissatisfying, partially because the compiler inevitably ends up getting blamed, but also because it means that huge bodies of C code are land mines just waiting to explode. This is even worse because [...] there is no good way to determine whether a large scale application is free of undefined behavior, and thus not susceptible to breaking in the future. 
\end{quote}
There is a proposal \cite{eskil}  in front of the ISO C Standards committee (WG14) to curtail undefined behavior semantics, but it is controversial.

 \subsection{Arithmetic Overflow}
An example of how undefined behavior works in practice for arithmetic overflow was explained by Lattner \cite{lattner}.
\begin{quote}
    knowing that INT\_MAX+1 is undefined allows optimizing X+1 > X to “true”. Knowing the multiplication “cannot” overflow (because doing so would be undefined) allows optimizing X*2/2 to X.  
\end{quote}
C "ints" are  fixed-size sequences of bytes interpreted as 2s complement values  that map into the ring \(\mathbb{Z}/2^k\mathbb{Z}\) where
\((x*y)/y = x\) is not a theorem\footnote{ Taking both \(x\leq INT\_MAX\) and \(x+1 >x\) as axioms implies that \(INT\_MAX > INT\_MAX\).}.
\textsf{Gcc} x86-64 with the optimizer on will reveal that (see the code and compilation \cite{calgebra1})  
if \(x\)  is an "int"
and \(x = 1,000,000,000\) then calculating \((x*5)/5\) directly produces \(1,000,000,000\) but
also 
\(z = x*5 = 705032704\)  and then  \(z/5 = 141006540\). The result 
depends on whether the compiler can recognize the overflows. Paradoxical results
are easy to generate.

Operating system programmers in Linux discovered this issue around  2007 when they found C code of the form \(if(index+length < index)\{...\}\) was being  silently deleted by the compiler (since it has to be false axiomatically), causing security and
logical failures \cite{felix}. 
Eventually, the  operating system (and other projects such as the Postgres database) resorted to a compiler-specific flag to force "wrapping semantics" outside of ISO C. The same interpretation justifies "optimizing" 
\begin{verbatim}
    while (i++ >= i) { adjust_valve(); }; 
    if(pressure_too_high())emergency();
\end{verbatim}
into an infinite loop that never gets to the "if" statement. 

There are many complex interactions between the "can't happen" interpretation of undefined behavior and C's rules for arithmetic and variable promotion. For example, modular arithmetic  is required for \emph{unsigned } arithmetic, but if "x" and "y" are unsigned short, and "z" is unsigned int, then the expression "z = x*y" can sometimes trigger undefined behavior. C "promotes" the two variables on the right to type "int"  in order to not lose precision, but then  signed integer arithmetic "can't overflow" and the compiler may sometimes assume, incorrectly,  the result is less than \verb|INT_MAX| \cite{eskilpromote};

\subsection{What is lost}

Choosing to maximise freedom for the compiler, while still specifying the language in a precise way, tends to increase the burden of complexity shared by implementors and users. Consider these optimization-related issues.

\subsubsection{Pointer casts}
According to the text of the C18 ISO C standard,
pointers of most types can be freely cast to other pointer types (section 6.3.2.3 paragraph 7).
\begin{quote}
    A pointer to an object type may be converted to a pointer to a different object type. If the resulting
pointer is not correctly aligned 69) for the referenced type, the behavior is undefined.
\end{quote}
However, complex rules govern what accesses are permitted via such pointers.
The pointer is correctly aligned in this code: \begin{verbatim}
float *f = malloc(sizeof(float));
*f = 3.14; //1
int *a =  (int *)f
*a = 4; //2
\end{verbatim} 
The language in 
section 6.5 paragraph 6 then implies that the "effective type" of an allocated object can be changed by writing to it:
\begin{quote}
If a value is stored into an object having no declared type through an lvalue having a type
that is not a character type, then the type of the lvalue becomes the effective type of the object for
that access and for subsequent accesses that do not modify the stored value.
\end{quote}
So have we converted the object pointed to by \verb|f|
to an int in statement 2?   Section  
6.5 paragraph 7 says \emph{"
An object shall have its stored value accessed only by an lvalue expression that has one of the following types"},
which are  
limited to compatible types and character types. We
can convert \verb|f| to be a pointer to an int, which
is even correctly aligned, but "access" includes both reads and writes.  
The assignment of statement 2 is (currently) compiled as written by both \textsf{gcc} and \textsf{Clang}, although the similar floating-point
example above omits the assignment to the aliasing pointer.
The rules here were substantially revised for C99,
but are still not considered adequate (e.g.\ even by many in WG14's Memory Object Model study group).
In C89 this same 
section allows access only by lvalues that have compatible \emph{declared} types, which appears to prevent any
access at all to allocated objects coming from malloc, so
statement 1 would be undefined behavior in C89 -- apparently an error in the specification.
Derek Jones \cite{derekjones} points out other changes in the C99 standard were required to allow memcpy to be written in C, following mistakes in C89 that prevented this.

Related issues include the behaviour of freed pointers \cite{pointerzap}, comparison of `one-past-the-end' pointers \cite{jmyers}, and semantics of integer-to-pointer casts \cite{kang}. The latter proposes a `quasi-concrete' semantics, where casting a pointer to an integer limits the permitted optimizations. This restraint is relatively rare -- it is unclear whether its compromise will be acceptable to ISO -- yet still misses some cases where addresses may be validly known to the wider program.

\subsubsection{Temporally unbounded UB}
The WG14 Memory Object Model study group was started in order to come up with a proposal for memory and pointer semantics. Their working proposal \cite{provenance}
is quoted twice above, but also explains: 
 \begin{quote} For evaluation-order and concurrency nondeterminism, one
would normally say that if there exists any execution that flags UB, then the program as a whole has UB (for the
moment ignoring UB that occurs only on some paths following I/O input, which is another important question
that the current ISO text does not address).
This view of UB seems to be unfortunate but inescapable. If one looks just at a single execution, then (at least
between input points) we cannot temporally bound the effects of an UB, because compilers can and do re-order
code w.r.t. the C abstract machine’s sequencing of computation. In other words, UB may be flagged at some
specific point in an abstract-machine trace, but its consequences on the observed implementation behaviour might
happen much earlier (in practice, perhaps not very much earlier, but we do not have any good way of bounding how
much). But then if one execution might have UB, and hence exhibit (in an implementation) arbitrary observable
behaviour, then anything the standard might say about any other execution is irrelevant, because it can always
be masked by that arbitrary observable behaviour.
\end{quote}

Perhaps the unifying theme of both of these topics can be found in Dennis Ritchie's comment on noalias cited above:\emph{"
 I don't know any acceptable way of changing the language specification to express the possibility of this kind of optimization"}. 
 C is stubbornly low-level and changing the language specification to permit these types of optimizations is hard, or maybe impossible. Current proposals introduce highly complex rules, which despite their complexity are known to be inadequate for certain systems programming idioms. If accepted by ISO, they are likely to be misunderstood by both practitioners and implementors, perpetuating rather than solving the problem. As one possible solution, Torvalds  \cite{torvaldsmm} and Ertl  \cite{ertl2} both propose relatively concrete, operational alternatives -- where compilers
map source operations to well-defined instruction sequences, in either a
virtual or real machine, from which compiler  optimisations may not
observably stray.

\section{What is gained}
The second part of Ritchie's comment cited above is \emph{"and I don't know how much performance improvement is likely to result."} 
It is difficult to find any documentation of significant performance advantages of 
any kind of undefined behavior optimization \cite{ertlwhat}. 
The standard TBAA lifting example can be better and more generally optimized by hand without
much effort -- without needing a type mismatch. Wang \emph{et al} ( \cite{wang}, Sec. 3.3) could not find a case where a UB optimization could not be 
matched with simple coding changes. Other optimizations are also mostly justified by small differences in SPEC benchmarks
or references to proprietary data sets.

Lee \cite{lee} provides an example of the claimed advantage of undefined behavior for 
overflow.
\begin{verbatim}
    for(int i=0 ; i <= N; i++)a[i] = x+1;
\end{verbatim}
If \(i\) and \(N\) are both 32bit and the target machine is x86-64, it is sometimes assumed
that permitting overflow requires a sign extend
of \verb|i| on each iteration. 
\begin{verbatim}
.L3:    movsx   rcx, eax #64 bit sign extend i
        add     eax, 1
        mov     DWORD PTR [rsi+rcx*4], edx
        cmp     edi, eax
        jge     .L3
\end{verbatim}
Assuming overflow is impossible allows omitting the sign extend.
Sign extension is a fast operation so if the loop
is non-trivial, the cost will be lost in the noise. In any case, 
\verb|i| can only overflow when \verb|N == INT_MAX|. The code then
executes an infinite loop, writing \verb|x+1| from \(a[INT\_MIN]\) to \(a[INT\_MAX]\).
\textsf{Gcc} will omit the sign extend all the same, if
the programmer replaces \verb|<=| with \verb|<|.
In sum,  the "slowdown" is caused by a single case that is nearly certainly an error and easily avoided.
As with many other similar cases, this example of ISO C optimizations turns out to to depend 
on the assumption that C programmers will not profile and optimize their own code. 

Even for aliasing-based optimizations, 
it is not necessary for alias analysis to depend on undefined behavior. 
 Alias detection in the abstract is not Turing computable \cite{aliasundecid} and C pointers make approximate alias detection difficult \cite{alias2}, but there are effective algorithms that can detect most aliasing \cite{Hind2000WhichPA} and it is a design 
 choice to make aliasing optimization rely on assumptions about program code that are not validated. ISO C has chosen
 to reduce the burden on the compilers at the expense of semantic clarity. 

The CompCert compiler is aimed at control systems that 
have many of the same properties as operating systems, does not do any undefined behavior based
optimization\footnote{Except for assuming objects do not overlap in memory.} (and does not optimize extensively) \cite{leroypersonal} and has a deterministic semantics \cite{Leroy-Compcert-CACM}:
\begin{quote}
    The semantics is deterministic and makes
precise a number of behaviors left unspecified or undefined
in the ISO C standard
[...]

CompCert generates code
that is more than twice as fast as that generated by \textsf{gcc}
without optimizations, and competitive with \textsf{gcc} at optimization levels 1 and 2. On average, CompCert code is
only 7\% slower than \textsf{gcc} -O1 and 12\% slower than \textsf{gcc} -O2.
\end{quote}

Finally, there is an experiment done at RedHat by Vladimir Makarov \cite{vlad}:
\begin{quote}
 I did an experiment by switching on only a fast and simple RA and combiner in \textsf{gcc}. There are no options to do this, I needed to modify \textsf{gcc}. [..] Compared to hundreds of optimizations in \textsf{gcc}-9.0 with -O2, these two optimizations achieve almost 80\% performance on an Intel i7-9700K machine under Fedora Core 29 for real-world programs through SpecCPU, one of the most credible compiler benchmarks.
\end{quote}
Makarov then tested a 20 year old version of \textsf{gcc} on 
Spec benchmarks versus a contemporary version of the compiler to show a 16\%  improvement with all
optimizations enabled --- over a period where most UB based optimizations went into the compilers.

A small  performance improvement will generally not justify a decrease in code
stability for operating systems or avionics controllers, but the answer may be different for "at-scale" data center applications or large
numerical simulations. 

\appendix

\section{Acknowledgments} 
 The perceptive comments and suggestions of the PLOS reviewers and paper shepherd, Stephen Kell, gratefully acknowledged. 
  Thanks to John Regehr, Eskil Steenberg, Anton Ertl, 
Paul E. McKenney, Rich Felker, and members of SC22/WG14 and the Memory Object Model discussion group for education, 
discussion, and disagreements. I am responsible for any errors. 
\bibliographystyle{ACM-Reference-Format}
\bibliography{base}


\begin{thebibliography}{44}


\ifx \showCODEN    \undefined \def \showCODEN     #1{\unskip}     \fi
\ifx \showDOI      \undefined \def \showDOI       #1{#1}\fi
\ifx \showISBNx    \undefined \def \showISBNx     #1{\unskip}     \fi
\ifx \showISBNxiii \undefined \def \showISBNxiii  #1{\unskip}     \fi
\ifx \showISSN     \undefined \def \showISSN      #1{\unskip}     \fi
\ifx \showLCCN     \undefined \def \showLCCN      #1{\unskip}     \fi
\ifx \shownote     \undefined \def \shownote      #1{#1}          \fi
\ifx \showarticletitle \undefined \def \showarticletitle #1{#1}   \fi
\ifx \showURL      \undefined \def \showURL       {\relax}        \fi
\providecommand\bibfield[2]{#2}
\providecommand\bibinfo[2]{#2}
\providecommand\natexlab[1]{#1}
\providecommand\showeprint[2][]{arXiv:#2}

\bibitem[\protect\citeauthoryear{Dietz, Li, Regehr, and Adve}{Dietz
  et~al\mbox{.}}{2012}]%
        {dietz}
\bibfield{author}{\bibinfo{person}{Will Dietz}, \bibinfo{person}{Peng Li},
  \bibinfo{person}{John Regehr}, {and} \bibinfo{person}{Vikram Adve}.}
  \bibinfo{year}{2012}\natexlab{}.
\newblock \showarticletitle{Understanding Integer Overflow in C/C++}.
\newblock \bibinfo{journal}{\emph{Proceedings - International Conference on
  Software Engineering}}  \bibinfo{volume}{25} (\bibinfo{date}{07}
  \bibinfo{year}{2012}).
\newblock
\urldef\tempurl%
\url{https://doi.org/10.1109/ICSE.2012.6227142}
\showDOI{\tempurl}


\bibitem[\protect\citeauthoryear{Ertl}{Ertl}{2015}]%
        {ertlwhat}
\bibfield{author}{\bibinfo{person}{M.~Anton Ertl}.}
  \bibinfo{year}{2015}\natexlab{}.
\newblock \showarticletitle{What every compiler writer should know about
  programmers}. In \bibinfo{booktitle}{\emph{18. Kolloquium Programmiersprachen
  und Grundlagen der Programmierung (KPS'15)}},
  \bibfield{editor}{\bibinfo{person}{Jens Knoop} {and}
  \bibinfo{person}{M.~Anton Ertl}} (Eds.). \bibinfo{pages}{112--133}.
\newblock
\urldef\tempurl%
\url{http://www.complang.tuwien.ac.at/kps2015/proceedings/KPS_2015_submission_29.pdf}
\showURL{%
\tempurl}


\bibitem[\protect\citeauthoryear{Ertl}{Ertl}{2017}]%
        {ertl2}
\bibfield{author}{\bibinfo{person}{M.~Anton Ertl}.}
  \bibinfo{year}{2017}\natexlab{}.
\newblock \showarticletitle{The Intended Meaning of \emph{Undefined Behaviour}
  in {C} Programs}. In \bibinfo{booktitle}{\emph{19. Kolloquium
  Programmiersprachen und Grundlagen der Programmierung (KPS'17)}},
  \bibfield{editor}{\bibinfo{person}{Wolfram Amme} {and}
  \bibinfo{person}{Thomas Heinze}} (Eds.). \bibinfo{pages}{20--28}.
\newblock
\urldef\tempurl%
\url{http://www.complang.tuwien.ac.at/papers/ertl17kps.pdf}
\showURL{%
\tempurl}


\bibitem[\protect\citeauthoryear{Felix-gcc}{Felix-gcc}{2007}]%
        {felix}
\bibfield{author}{\bibinfo{person}{Felix-gcc}.}
  \bibinfo{year}{2007}\natexlab{}.
\newblock \bibinfo{title}{Bug 30475 - assert(int+100 > int) optimized away}.
\newblock
\newblock
\urldef\tempurl%
\url{https://gcc.gnu.org/bugzilla/show_bug.cgi?id=30475}
\showURL{%
\tempurl}


\bibitem[\protect\citeauthoryear{Gustedt, Sewell, Memarian, Gomes, and
  Uecker}{Gustedt et~al\mbox{.}}{2021}]%
        {provenance}
\bibfield{author}{\bibinfo{person}{Jens Gustedt}, \bibinfo{person}{Peter
  Sewell}, \bibinfo{person}{Kayvan Memarian}, \bibinfo{person}{Victor B.~F.
  Gomes}, {and} \bibinfo{person}{Martin Uecker}.}
  \bibinfo{year}{2021}\natexlab{}.
\newblock \bibinfo{title}{A Provenance-aware Memory Object Model for C. Draft
  Technical Specification N2577}.
\newblock
\newblock
\urldef\tempurl%
\url{http://www.open-std.org/jtc1/sc22/wg14/www/docs/n2577.pdf}
\showURL{%
\tempurl}


\bibitem[\protect\citeauthoryear{Hathhorn, Ellison, and Ro\c{s}u}{Hathhorn
  et~al\mbox{.}}{2015}]%
        {hathorn}
\bibfield{author}{\bibinfo{person}{Chris Hathhorn}, \bibinfo{person}{Chucky
  Ellison}, {and} \bibinfo{person}{Grigore Ro\c{s}u}.}
  \bibinfo{year}{2015}\natexlab{}.
\newblock \showarticletitle{Defining the Undefinedness of C}. In
  \bibinfo{booktitle}{\emph{Proceedings of the 36th ACM SIGPLAN Conference on
  Programming Language Design and Implementation}} (Portland, OR, USA)
  \emph{(\bibinfo{series}{PLDI '15})}. \bibinfo{publisher}{Association for
  Computing Machinery}, \bibinfo{address}{New York, NY, USA},
  \bibinfo{pages}{336–345}.
\newblock
\showISBNx{9781450334686}
\urldef\tempurl%
\url{https://doi.org/10.1145/2737924.2737979}
\showDOI{\tempurl}


\bibitem[\protect\citeauthoryear{Hind and Pioli}{Hind and Pioli}{2000}]%
        {Hind2000WhichPA}
\bibfield{author}{\bibinfo{person}{M. Hind} {and} \bibinfo{person}{Anthony
  Pioli}.} \bibinfo{year}{2000}\natexlab{}.
\newblock \showarticletitle{Which pointer analysis should I use?}. In
  \bibinfo{booktitle}{\emph{ISSTA '00}}.
\newblock


\bibitem[\protect\citeauthoryear{{ISO PL22.11 - SC22/WG14 }}{{ISO PL22.11 -
  SC22/WG14 }}{2018}]%
        {c18}
\bibfield{author}{\bibinfo{person}{{ISO PL22.11 - SC22/WG14 }}.}
  \bibinfo{year}{2018}\natexlab{}.
\newblock \bibinfo{booktitle}{\emph{Programming language: {C}: ISO/IEC
  9899:2018 (C18)}}.
\newblock Number ISO/IEC 9899:2018).
\newblock


\bibitem[\protect\citeauthoryear{J11 and WG14}{J11 and WG14}{2003}]%
        {rationalec}
\bibfield{author}{\bibinfo{person}{INCITS J11} {and} \bibinfo{person}{SC22
  WG14}.} \bibinfo{year}{2003}\natexlab{}.
\newblock \bibinfo{title}{Rationale for International Standard. Programming
  Languages. C Revision 5.10}.
\newblock
\newblock
\urldef\tempurl%
\url{http://www.open-std.org/jtc1/sc22/wg14/www/C99RationaleV5.10.pdf}
\showURL{%
\tempurl}


\bibitem[\protect\citeauthoryear{Jones}{Jones}{2017}]%
        {derekjones}
\bibfield{author}{\bibinfo{person}{Derek Jones}.}
  \bibinfo{year}{2017}\natexlab{}.
\newblock \bibinfo{title}{How indeterminate is an indeterminate value}.
\newblock
\newblock
\urldef\tempurl%
\url{http://shape-of-code.coding-guidelines.com/2017/06/18/how-indeterminate-is-an-indeterminate-value/}
\showURL{%
\tempurl}


\bibitem[\protect\citeauthoryear{Kang, Hur, Mansky, Garbuzov, Zdancewic, and
  Vafeiadis}{Kang et~al\mbox{.}}{2015}]%
        {kang}
\bibfield{author}{\bibinfo{person}{Jeehoon Kang}, \bibinfo{person}{Chung-Kil
  Hur}, \bibinfo{person}{William Mansky}, \bibinfo{person}{Dmitri Garbuzov},
  \bibinfo{person}{Steve Zdancewic}, {and} \bibinfo{person}{Viktor Vafeiadis}.}
  \bibinfo{year}{2015}\natexlab{}.
\newblock \showarticletitle{A Formal C Memory Model Supporting Integer-Pointer
  Casts}.
\newblock \bibinfo{journal}{\emph{SIGPLAN Not.}} \bibinfo{volume}{50},
  \bibinfo{number}{6} (\bibinfo{date}{June} \bibinfo{year}{2015}),
  \bibinfo{pages}{326–335}.
\newblock
\showISSN{0362-1340}
\urldef\tempurl%
\url{https://doi.org/10.1145/2813885.2738005}
\showDOI{\tempurl}


\bibitem[\protect\citeauthoryear{Kernighan}{Kernighan}{[n.d.]}]%
        {kernighan}
\bibfield{author}{\bibinfo{person}{Brian~W. Kernighan}.}
  \bibinfo{year}{[n.d.]}\natexlab{}.
\newblock \bibinfo{title}{Why Pascal is Not My Favorite Programming Language}.
\newblock
\newblock
\urldef\tempurl%
\url{http://www.lysator.liu.se/c/bwk-on-pascal.html}
\showURL{%
\tempurl}


\bibitem[\protect\citeauthoryear{Kernighan and Ritchie}{Kernighan and
  Ritchie}{1988}]%
        {kandr}
\bibfield{author}{\bibinfo{person}{Brian~W. Kernighan} {and}
  \bibinfo{person}{Dennis~M. Ritchie}.} \bibinfo{year}{1988}\natexlab{}.
\newblock \bibinfo{booktitle}{\emph{The C Programming Language}
  (\bibinfo{edition}{2nd} ed.)}.
\newblock \bibinfo{publisher}{Prentice Hall Professional Technical Reference}.
\newblock
\showISBNx{0131103709}


\bibitem[\protect\citeauthoryear{Landi and Ryder}{Landi and Ryder}{1992}]%
        {alias2}
\bibfield{author}{\bibinfo{person}{W. Landi} {and} \bibinfo{person}{B. Ryder}.}
  \bibinfo{year}{1992}\natexlab{}.
\newblock \showarticletitle{A safe approximate algorithm for interprocedural
  aliasing}. In \bibinfo{booktitle}{\emph{PLDI '92}}.
\newblock


\bibitem[\protect\citeauthoryear{Lattner}{Lattner}{2011a}]%
        {lattner}
\bibfield{author}{\bibinfo{person}{Chris Lattner}.}
  \bibinfo{year}{2011}\natexlab{a}.
\newblock \bibinfo{title}{What every C programmer should know}.
\newblock
\newblock
\urldef\tempurl%
\url{http://blog.llvm.org/2011/05/what-every-c-programmer-should-know.html}
\showURL{%
\tempurl}


\bibitem[\protect\citeauthoryear{Lattner}{Lattner}{2011b}]%
        {lattner2}
\bibfield{author}{\bibinfo{person}{Chris Lattner}.}
  \bibinfo{year}{2011}\natexlab{b}.
\newblock \bibinfo{title}{What Every C Programmer Should Know About Undefined
  Behavior 2/3}.
\newblock
\newblock
\urldef\tempurl%
\url{https://blog.llvm.org/2011/05/what-every-c-programmer-should-know_14.html}
\showURL{%
\tempurl}


\bibitem[\protect\citeauthoryear{Lee, Kim, Song, Hur, Das, Majnemer, Regehr,
  and Lopes}{Lee et~al\mbox{.}}{2017}]%
        {lee}
\bibfield{author}{\bibinfo{person}{Juneyoung Lee}, \bibinfo{person}{Yoonseung
  Kim}, \bibinfo{person}{Youngju Song}, \bibinfo{person}{Chung-Kil Hur},
  \bibinfo{person}{Sanjoy Das}, \bibinfo{person}{David Majnemer},
  \bibinfo{person}{John Regehr}, {and} \bibinfo{person}{Nuno~P. Lopes}.}
  \bibinfo{year}{2017}\natexlab{}.
\newblock \showarticletitle{Taming Undefined Behavior in LLVM}.
\newblock \bibinfo{journal}{\emph{SIGPLAN Not.}} \bibinfo{volume}{52},
  \bibinfo{number}{6} (\bibinfo{date}{June} \bibinfo{year}{2017}),
  \bibinfo{pages}{633–647}.
\newblock
\showISSN{0362-1340}
\urldef\tempurl%
\url{https://doi.org/10.1145/3140587.3062343}
\showDOI{\tempurl}


\bibitem[\protect\citeauthoryear{Leroy}{Leroy}{2009}]%
        {Leroy-Compcert-CACM}
\bibfield{author}{\bibinfo{person}{Xavier Leroy}.}
  \bibinfo{year}{2009}\natexlab{}.
\newblock \showarticletitle{Formal verification of a realistic compiler}.
\newblock \bibinfo{journal}{\emph{Commun. ACM}} \bibinfo{volume}{52},
  \bibinfo{number}{7} (\bibinfo{year}{2009}), \bibinfo{pages}{107--115}.
\newblock
\urldef\tempurl%
\url{http://xavierleroy.org/publi/compcert-CACM.pdf}
\showURL{%
\tempurl}


\bibitem[\protect\citeauthoryear{Leroy}{Leroy}{2021}]%
        {leroypersonal}
\bibfield{author}{\bibinfo{person}{Xavier Leroy}.}
  \bibinfo{year}{2021}\natexlab{}.
\newblock \bibinfo{title}{Personal Communication}.
\newblock
\newblock


\bibitem[\protect\citeauthoryear{Leroy, Appel, Blazy, and Stewart}{Leroy
  et~al\mbox{.}}{2012}]%
        {compcert2}
\bibfield{author}{\bibinfo{person}{Xavier Leroy}, \bibinfo{person}{Andrew~W.
  Appel}, \bibinfo{person}{Sandrine Blazy}, {and} \bibinfo{person}{Gordon
  Stewart}.} \bibinfo{year}{2012}\natexlab{}.
\newblock \bibinfo{booktitle}{\emph{{The CompCert Memory Model, Version 2}}}.
\newblock \bibinfo{type}{Research Report} RR-7987.
  \bibinfo{institution}{{INRIA}}. \bibinfo{pages}{26} pages.
\newblock
\urldef\tempurl%
\url{https://hal.inria.fr/hal-00703441}
\showURL{%
\tempurl}


\bibitem[\protect\citeauthoryear{Makarov}{Makarov}{2020}]%
        {vlad}
\bibfield{author}{\bibinfo{person}{Vladimir Makarov}.}
  \bibinfo{year}{2020}\natexlab{}.
\newblock \bibinfo{title}{MIR: A lightweight JIT compiler project}.
\newblock
\newblock
\urldef\tempurl%
\url{https://developers.redhat.com/blog/2020/01/20/mir-a-lightweight-jit-compiler-project}
\showURL{%
\tempurl}


\bibitem[\protect\citeauthoryear{McKenney, Michael, Mauer, Sewell, Uecker,
  Boehm, Tong, and Douglas}{McKenney et~al\mbox{.}}{2019}]%
        {pointerzap}
\bibfield{author}{\bibinfo{person}{Paul~E. McKenney}, \bibinfo{person}{Maged
  Michael}, \bibinfo{person}{Jens Mauer}, \bibinfo{person}{Peter Sewell},
  \bibinfo{person}{Martin Uecker}, \bibinfo{person}{Hans Boehm},
  \bibinfo{person}{Hubert Tong}, {and} \bibinfo{person}{Niall Douglas}.}
  \bibinfo{year}{2019}\natexlab{}.
\newblock \bibinfo{title}{Pointer lifetime-end zap}.
\newblock
\newblock
\urldef\tempurl%
\url{http://www.open-std.org/jtc1/sc22/wg21/docs/papers/2019/p1726r0.pdf}
\showURL{%
\tempurl}


\bibitem[\protect\citeauthoryear{Myers}{Myers}{2014}]%
        {jmyers}
\bibfield{author}{\bibinfo{person}{Joseph Myers}.}
  \bibinfo{year}{2014}\natexlab{}.
\newblock \bibinfo{title}{"Bug 61502: comparison on "one-past" pointer gives
  wrong result, comment 1"}.
\newblock
\newblock
\urldef\tempurl%
\url{https://gcc.gnu.org/bugzilla/show_bug.cgi?id=61502}
\showURL{%
\tempurl}


\bibitem[\protect\citeauthoryear{Nguyen and Irigoin}{Nguyen and
  Irigoin}{2003}]%
        {nguyenalias}
\bibfield{author}{\bibinfo{person}{Thi Viet~Nga Nguyen} {and}
  \bibinfo{person}{Fran{\c{c}}ois Irigoin}.} \bibinfo{year}{2003}\natexlab{}.
\newblock \showarticletitle{Alias verification for Fortran code optimization}.
\newblock \bibinfo{journal}{\emph{J. UCS}} \bibinfo{volume}{9},
  \bibinfo{number}{3} (\bibinfo{year}{2003}), \bibinfo{pages}{270}.
\newblock


\bibitem[\protect\citeauthoryear{Nossum}{Nossum}{2016}]%
        {nossum}
\bibfield{author}{\bibinfo{person}{Vegard Nossum}.}
  \bibinfo{year}{2016}\natexlab{}.
\newblock \bibinfo{title}{"Subject [PATCH] firmware: declare
  \_\_{start,end}\_builtin\_fw as pointers"}.
\newblock
\newblock
\urldef\tempurl%
\url{https://gcc.gnu.org/bugzilla/show_bug.cgi?id=61502}
\showURL{%
\tempurl}


\bibitem[\protect\citeauthoryear{on~the C~Programming~Language}{on~the
  C~Programming~Language}{1989}]%
        {ANSIC89}
\bibfield{author}{\bibinfo{person}{X3J11 Technical~Committee on~the
  C~Programming~Language}.} \bibinfo{year}{1989}\natexlab{}.
\newblock \bibinfo{title}{ANSI X3.159, 1989 Edition, 1989 - INFORMATION SYSTEMS
  - PROGRAMMING LANGUAGE - C}.
\newblock
\newblock


\bibitem[\protect\citeauthoryear{Project}{Project}{2021a}]%
        {clang}
\bibfield{author}{\bibinfo{person}{Clang Project}.}
  \bibinfo{year}{2021}\natexlab{a}.
\newblock \bibinfo{title}{Clang 13 Documentation}.
\newblock
\newblock
\urldef\tempurl%
\url{https://clang.llvm.org/docs/UsersManual.html}
\showURL{%
\tempurl}


\bibitem[\protect\citeauthoryear{Project}{Project}{2021b}]%
        {gcc}
\bibfield{author}{\bibinfo{person}{GNU Project}.}
  \bibinfo{year}{2021}\natexlab{b}.
\newblock \bibinfo{title}{GCC Documentation}.
\newblock
\newblock
\urldef\tempurl%
\url{https://gcc.gnu.org/onlinedocs/}
\showURL{%
\tempurl}


\bibitem[\protect\citeauthoryear{Ramalingam}{Ramalingam}{1994}]%
        {aliasundecid}
\bibfield{author}{\bibinfo{person}{Ganesan Ramalingam}.}
  \bibinfo{year}{1994}\natexlab{}.
\newblock \showarticletitle{The undecidability of aliasing}.
\newblock \bibinfo{journal}{\emph{ACM Transactions on Programming Languages and
  Systems (TOPLAS)}} \bibinfo{volume}{16}, \bibinfo{number}{5}
  (\bibinfo{year}{1994}), \bibinfo{pages}{1467--1471}.
\newblock


\bibitem[\protect\citeauthoryear{Regehr}{Regehr}{2010}]%
        {regehrguide}
\bibfield{author}{\bibinfo{person}{John Regehr}.}
  \bibinfo{year}{2010}\natexlab{}.
\newblock
\newblock
\urldef\tempurl%
\url{https://blog.regehr.org/archives/213}
\showURL{%
\tempurl}


\bibitem[\protect\citeauthoryear{Regehr}{Regehr}{2016}]%
        {regehralias}
\bibfield{author}{\bibinfo{person}{John Regehr}.}
  \bibinfo{year}{2016}\natexlab{}.
\newblock \bibinfo{title}{The Strict Aliasing Situation is Pretty Bad}.
\newblock
\newblock
\urldef\tempurl%
\url{https://blog.regehr.org/archives/1307}
\showURL{%
\tempurl}


\bibitem[\protect\citeauthoryear{Ritchie}{Ritchie}{1988}]%
        {dmr}
\bibfield{author}{\bibinfo{person}{Dennis Ritchie}.}
  \bibinfo{year}{1988}\natexlab{}.
\newblock \showarticletitle{noalias comments to X3J11}.
\newblock  (\bibinfo{date}{March} \bibinfo{year}{1988}).
\newblock
\urldef\tempurl%
\url{https://groups.google.com/g/comp.lang.c/c/K0Cz2s9il3E/m/YDyo_xaRG5kJ}
\showURL{%
\tempurl}


\bibitem[\protect\citeauthoryear{Ritchie, Johnson, Lesk, and Kernighan}{Ritchie
  et~al\mbox{.}}{1978}]%
        {ritchie1978c}
\bibfield{author}{\bibinfo{person}{DM Ritchie}, \bibinfo{person}{SC Johnson},
  \bibinfo{person}{ME Lesk}, {and} \bibinfo{person}{BW Kernighan}.}
  \bibinfo{year}{1978}\natexlab{}.
\newblock \showarticletitle{The C programming language, Bell Systems Tech}.
\newblock \bibinfo{journal}{\emph{J}} \bibinfo{volume}{57}, \bibinfo{number}{6}
  (\bibinfo{year}{1978}), \bibinfo{pages}{1991--2020}.
\newblock


\bibitem[\protect\citeauthoryear{Steenberg}{Steenberg}{2021a}]%
        {eskilpromote}
\bibfield{author}{\bibinfo{person}{Eskil Steenberg}.}
  \bibinfo{year}{2021}\natexlab{a}.
\newblock \bibinfo{title}{"Compiler Explorer UShort promotion UB"}.
\newblock
\newblock
\urldef\tempurl%
\url{https://godbolt.org/z/7q9dPzEfM}
\showURL{%
\tempurl}


\bibitem[\protect\citeauthoryear{Steenberg}{Steenberg}{2021b}]%
        {eskil}
\bibfield{author}{\bibinfo{person}{Eskil Steenberg}.}
  \bibinfo{year}{2021}\natexlab{b}.
\newblock \showarticletitle{Redefining Undefined Behavior N2769}.
\newblock  (\bibinfo{date}{21 6} \bibinfo{year}{2021}).
\newblock
\urldef\tempurl%
\url{http://www.open-std.org/jtc1/sc22/wg14/www/docs/n2769.pdf}
\showURL{%
\tempurl}


\bibitem[\protect\citeauthoryear{Torvalds}{Torvalds}{[n.d.]}]%
        {relochide}
\bibfield{author}{\bibinfo{person}{Linus Torvalds}.}
  \bibinfo{year}{[n.d.]}\natexlab{}.
\newblock \bibinfo{title}{Reloc-Hide in Linux Kernel}.
\newblock
\newblock
\urldef\tempurl%
\url{https://github.com/torvalds/linux/blob/35e43538af8fd2cb39d58caca1134a87db173f75/include/linux/compiler-gcc.h}
\showURL{%
\tempurl}


\bibitem[\protect\citeauthoryear{Torvalds}{Torvalds}{2009}]%
        {torvaldsalias2}
\bibfield{author}{\bibinfo{person}{Linus Torvalds}.}
  \bibinfo{year}{2009}\natexlab{}.
\newblock \bibinfo{title}{Re Gcc inlining heuristics}.
\newblock
\newblock
\urldef\tempurl%
\url{https://www.mail-archive.com/linux-btrfs@vger.kernel.org/msg01647.html}
\showURL{%
\tempurl}


\bibitem[\protect\citeauthoryear{Torvalds}{Torvalds}{2018a}]%
        {torvaldsalias}
\bibfield{author}{\bibinfo{person}{Linus Torvalds}.}
  \bibinfo{year}{2018}\natexlab{a}.
\newblock \bibinfo{title}{Re: [GIT PULL] Device properties framework update for
  v4.18-rc1}.
\newblock
\newblock
\urldef\tempurl%
\url{https://lkml.org/lkml/2018/6/5/769}
\showURL{%
\tempurl}


\bibitem[\protect\citeauthoryear{Torvalds}{Torvalds}{2018b}]%
        {torvaldsmm}
\bibfield{author}{\bibinfo{person}{Linus Torvalds}.}
  \bibinfo{year}{2018}\natexlab{b}.
\newblock \bibinfo{title}{Re: LKMM litmus test for Roman Penyaev's rcu-rr}.
\newblock
\newblock
\urldef\tempurl%
\url{https://lkml.org/lkml/2018/6/7/761}
\showURL{%
\tempurl}


\bibitem[\protect\citeauthoryear{Wang, Chen, Cheung, Jia, Zeldovich, and
  Kaashoek}{Wang et~al\mbox{.}}{2012}]%
        {wang}
\bibfield{author}{\bibinfo{person}{Xi Wang}, \bibinfo{person}{Haogang Chen},
  \bibinfo{person}{Alvin Cheung}, \bibinfo{person}{Zhihao Jia},
  \bibinfo{person}{Nickolai Zeldovich}, {and} \bibinfo{person}{M.~Frans
  Kaashoek}.} \bibinfo{year}{2012}\natexlab{}.
\newblock \showarticletitle{Undefined Behavior: What Happened to My Code?}. In
  \bibinfo{booktitle}{\emph{Proceedings of the Asia-Pacific Workshop on
  Systems}} (Seoul, Republic of Korea) \emph{(\bibinfo{series}{APSYS '12})}.
  \bibinfo{publisher}{Association for Computing Machinery},
  \bibinfo{address}{New York, NY, USA}, Article \bibinfo{articleno}{9},
  \bibinfo{numpages}{7}~pages.
\newblock
\showISBNx{9781450316699}
\urldef\tempurl%
\url{https://doi.org/10.1145/2349896.2349905}
\showURL{%
\tempurl}


\bibitem[\protect\citeauthoryear{Wang, Zeldovich, Kaashoek, and
  Solar-Lezama}{Wang et~al\mbox{.}}{2015}]%
        {wang2}
\bibfield{author}{\bibinfo{person}{Xi Wang}, \bibinfo{person}{Nickolai
  Zeldovich}, \bibinfo{person}{M.~Frans Kaashoek}, {and}
  \bibinfo{person}{Armando Solar-Lezama}.} \bibinfo{year}{2015}\natexlab{}.
\newblock \showarticletitle{A Differential Approach to Undefined Behavior
  Detection}.
\newblock \bibinfo{journal}{\emph{ACM Trans. Comput. Syst.}}
  \bibinfo{volume}{33}, \bibinfo{number}{1}, Article \bibinfo{articleno}{1}
  (\bibinfo{date}{March} \bibinfo{year}{2015}), \bibinfo{numpages}{29}~pages.
\newblock
\showISSN{0734-2071}
\urldef\tempurl%
\url{https://doi.org/10.1145/2699678}
\showDOI{\tempurl}


\bibitem[\protect\citeauthoryear{Wulf}{Wulf}{1972}]%
        {bliss}
\bibfield{author}{\bibinfo{person}{William~A Wulf}.}
  \bibinfo{year}{1972}\natexlab{}.
\newblock \showarticletitle{Systems for systems implementors: some experiences
  from Bliss}. In \bibinfo{booktitle}{\emph{Proceedings of the December 5-7,
  1972, fall joint computer conference, part II}}. \bibinfo{pages}{943--948}.
\newblock


\bibitem[\protect\citeauthoryear{Yodaiken}{Yodaiken}{2021a}]%
        {calgebra1}
\bibfield{author}{\bibinfo{person}{Victor Yodaiken}.}
  \bibinfo{year}{2021}\natexlab{a}.
\newblock \bibinfo{title}{Compiler Explorer ISO C Division}.
\newblock
\newblock
\urldef\tempurl%
\url{https://godbolt.org/z/zWh9c5e84}
\showURL{%
\tempurl}


\bibitem[\protect\citeauthoryear{Yodaiken}{Yodaiken}{2021b}]%
        {aliasfloat}
\bibfield{author}{\bibinfo{person}{Victor Yodaiken}.}
  \bibinfo{year}{2021}\natexlab{b}.
\newblock \bibinfo{title}{Example of Clang and type based alias}.
\newblock
\newblock
\urldef\tempurl%
\url{https://godbolt.org/z/nq19n8dhE}
\showURL{%
\tempurl}


\end{thebibliography}

\end{document}